\documentclass[11pt]{article}
\usepackage{cite}
\usepackage{scalerel}
\usepackage{shuffle}
\usepackage{amsmath,amsfonts,amssymb}
\usepackage{latexsym,epsfig}
\usepackage{dsfont}
\usepackage{hyperref}
\usepackage{extarrows}
\usepackage{comment} 
\usepackage{tikz-cd}

\def\hybrid{
        \topmargin -20pt
        \oddsidemargin 0pt
        \headheight 0pt \headsep 0pt
        \textwidth 6.25in 
        \textheight 9.5in 
        \marginparwidth .875in
        \parskip 5pt plus 1pt \jot = 1.5ex}

\hybrid

 \csname
@addtoreset\endcsname{equation}{section}

\def\cL{{\cal L}}
\def\cD{\mathcal{D}}

\def\cE{{\cal E}}

\def\cN{{\cal N}}
\def\cR{{\cal R}}

\def\cV{{\cal V}}

\def\cK{{\cal K}}

\def\cX{{\cal X}}

\def\del{\partial}
\def\l{\langle}
\def\r{\rangle}
\def\B{\square}

\allowdisplaybreaks

\def\bpm{\begin{pmatrix}}
\def\epm{\end{pmatrix}}

\begin{document}

\begin{titlepage}
\rightline{}
\rightline{April  2024}
\rightline{HU-EP-24/12-RTG}  
\begin{center}
\vskip 1.5cm
{\Large \bf{Double Copy of 3D Chern-Simons Theory \\[1.5ex] 
and 6D Kodaira-Spencer Gravity }}

\vskip 1.7cm

{\large\bf {Roberto Bonezzi, Felipe D\'iaz-Jaramillo and Olaf Hohm}}
\vskip 1.6cm

{\it  Institute for Physics, Humboldt University Berlin,\\
 Zum Gro\ss en Windkanal 6, D-12489 Berlin, Germany}\\[1.5ex] 
 ohohm@physik.hu-berlin.de, 
roberto.bonezzi@physik.hu-berlin.de, 
felipe.diaz-jaramillo@hu-berlin.de 
\vskip .1cm

\vskip .2cm

\end{center}

\bigskip\bigskip
\begin{center} 
\textbf{Abstract}

\end{center} 
\begin{quote}

We apply an algebraic  double copy construction of gravity from gauge theory to three-dimensional (3D) Chern-Simons 
theory. The kinematic algebra ${\cal K}$ is 
the 3D de Rham complex of forms equipped,  
for a choice of metric, with a graded Lie algebra 
that is equivalent to the Schouten-Nijenhuis bracket 
on polyvector fields. The double copied gravity is defined on a subspace of 
${\cal K}\otimes \bar{\cal K}$ and 
yields   a topological double field theory for  a generalized metric perturbation and 
two 2-forms. This local and gauge invariant theory is non-Lagrangian but  can be rendered Lagrangian 
by abandoning locality. Upon fixing a gauge this reduces to the double copy of Chern-Simons 
theory previously proposed by Ben-Shahar and Johansson. 
Furthermore, using  complex coordinates in $\mathbb{C}^3$ this  theory  is related 
to six-dimensional (6D)  Kodaira-Spencer gravity 
in that truncating the two 2-forms and one equation 
yields the Kodaira-Spencer equations on a 3D real slice of $\mathbb{C}^3$. 
The full 6D Kodaira-Spencer theory  can instead be obtained as a consistent truncation of a chiral double copy.

\end{quote} 
\vfill
\setcounter{footnote}{0}
\end{titlepage}

\tableofcontents

\section{Introduction} 

The double copy denotes modern amplitude techniques that relate 
the scattering amplitudes of gauge theory to those of gravity. The prime example 
is the double copy of pure Yang-Mills theory which yields at least at tree-level  Einstein gravity coupled 
to a B-field (two-form)  and a scalar (dilaton), a theory also known as  
`${\cal N}=0$ supergravity' \cite{Bern:2008qj,Bern:2019prr}. The double copy was originally defined at the 
level of scattering amplitudes and hence is a priori only meaningful for on-shell and gauge-fixed fields, 
but there are several reasons why one would like to go beyond this. 
For instance, one would like to double copy classical solutions of Yang-Mills theory to obtain classical gravity solutions, see \cite{Monteiro:2014cda, Luna:2015paa, Luna:2016hge, Luna:2018dpt, Kim:2019jwm, Monteiro:2021ztt, Monteiro:2020plf}, 
or one would like to have better control over double copy at loop level \cite{Bern:2007hh, Bern:2010ue, Carrasco:2011mn, Bjerrum-Bohr:2013iza, He:2015wgf, Bern:2018jmv, Edison:2022jln, Bern:2023zkg}.

In recent years there has been significant progress in this direction using the framework of homotopy algebras, 
based on the general dictionary between (classical) field theories and homotopy Lie or $L_{\infty}$ algebras \cite{Zwiebach:1992ie,Lada:1992wc,Zeitlin:2008cc,Hohm:2017pnh,Zeitlin:2009tj,Zeitlin:2014xma,Borsten:2021hua}. 
These are generalizations of differential graded Lie algebras, i.e., superalgebras equipped with a nilpotent 
operator, in which the Jacobi identity may only hold `up to homotopy' controlled by higher brackets. 
$L_{\infty}$ algebras naturally encode the data of a gauge field theory, including Yang-Mills theory and gravity, the latter 
written perturbatively as an expansion about a background metric.

Homotopy algebras are a powerful framework for double copy  since, among other reasons,  they allow one to 
give meaning to the notion of `stripping off' the color factors associated to the gauge group.  
Specifically, the $L_{\infty}$ algebra of Yang-Mills theory factorizes into the tensor product ${\cal X}_{\rm YM}={\cal K}\otimes \mathfrak{g}$, 
where $\mathfrak{g}$ is the Lie algebra of the gauge group and ${\cal K}$ the `kinematic' algebra,
which is a homotopy generalization of a differential graded commutative associative algebra or $C_{\infty}$ algebra \cite{Zeitlin:2008cc}. 
Importantly for double copy, the kinematic space ${\cal K}$ also carries hidden structures that define what has been termed a BV$_{\infty}^{\B}$ algebra \cite{Reiterer:2019dys, Borsten:2022vtg, Bonezzi:2022bse},  a generalization 
of a Batalin-Vilkovisky algebra \cite{Batalin:1981jr, CarrilloVallette}, where $\B$ denotes the wave operator.  
Upon taking the tensor product of two copies $\cK$ and $\bar\cK$ of the kinematic BV$_{\infty}^{\B}$ algebra, one can construct the $L_{\infty}$ algebra of ${\cal N}=0$ supergravity, formulated as a double field theory \cite{Hull:2009mi,Siegel:1993th,Hohm:2010pp}, on a suitable subspace of ${\cal K}\otimes \bar{\cal K}$. More precisely, so far this has been established to the order corresponding to quartic couplings in an action 
\cite{Bonezzi:2022bse,Bonezzi:2022yuh,Diaz-Jaramillo:2021wtl}.

In this note we apply the general algebraic double copy construction of \cite{Bonezzi:2022bse} to the toy model of three-dimensional (3D) 
Chern-Simons theory. This theory is topological, and hence all of its scattering amplitudes vanish, but it is still a 
fruitful toy model for color-kinematics duality (a necessary prerequisite for  the  double copy), 
which can be studied at the level of off-shell correlation functions \cite{Ben-Shahar:2021zww}. Here the `kinematic Lie algebra' underlying 
color-kinematics duality has been identified recently and just lives on the familiar de Rham complex of differential forms in 3D. 
This algebra is part of a BV$^{\B}$ algebra, which is \textit{strict}, meaning that no higher maps actually appear. 
One can thus straightforwardly apply the double copy construction of \cite{Bonezzi:2022bse} to 3D Chern-Simons theory and obtain 
a double copied 3D gravity theory in a formulation that is by construction local and gauge invariant. 
Here we spell out this 3D gravity theory. 
To the best of our knowledge this is the first  complete first-principle double copy construction of a diffeomorphism invariant 
gravity theory 
from a gauge theory. While the double copy of Chern-Simons theory has  been explored previously by 
Ben-Shahar and Johansson, their   gravity action is gauge fixed and non-local \cite{Ben-Shahar:2021zww}, see also  \cite{Szabo:2023cmv,Borsten:2023ned}. 
The double copied gravity theory constructed  in this paper is gauge invariant and local but non-Lagrangian. 
One can write down a gauge invariant  action, however, upon partial gauge fixing and abandoning  locality, 
which then reduces upon further gauge fixing and truncation 
to the action given in  \cite{Ben-Shahar:2021zww}.

Concretely, the 3D gravity theory obtained as the double copy of 3D Chern-Simons theory can be written 
in a form that exhibits a formal six-dimensional (6D) covariance as follows. The gauge field is a 6D 
2-form $\Psi= \frac{1}{2}\Psi_{MN}\theta^M \theta^N$, where  we view all differential forms as functions 
of even coordinates $x^M=(x^{\mu}, \bar{x}^{\bar{\mu}})$ and odd coordinates $\theta^M\simeq {\bf { d}}x^M$. 
The field equations  read 
 \begin{equation}\label{PSIEQINTRO} 
   {\bf d}\Psi + \tfrac{1}{2}\big[\Psi,\Psi\big] =0 \;, 
 \end{equation} 
which are invariant under gauge transformations with one-form parameters $\Lambda=\Lambda_M\theta^M$, 
 \begin{equation} 
  \delta \Psi = {\bf d} \Lambda + \big[\Psi,\Lambda\big]\, , 
 \end{equation} 
 where the graded symmetric bracket on general 6D forms is given by 
  \begin{equation}\label{SchoutenNiejenhuisINTRO} 
   \big[F_1, F_2\big] := \eta^{MN} \frac{\partial F_{1}}{\partial \theta^M} \frac{\partial F_{2}}{\partial x^N}\pm (1\leftrightarrow 2) \;, \qquad 
   \eta_{MN}=\bpm\delta_{\mu\nu}&0\\0&-\delta_{\bar\mu\bar\nu}\epm \;, 
  \end{equation} 
with the $O(3,3)$ invariant metric $\eta_{MN}$ built from two copies of a fiducial  3D metric indicated here by Kronecker deltas. 
More precisely, this theory is only gauge invariant and hence consistent when the fields are subject to the `strong constraint' 
making them effectively 3D. The reason is that while the bracket (\ref{SchoutenNiejenhuisINTRO}) defines a genuine graded  Lie algebra in 6D
(that written in terms of polyvector fields is known as the Schouten-Nijenhuis bracket), 
the 6D de Rham differential ${\bf d}=\theta^M\partial_M$ does not act via the Leibniz rule on the bracket. 
The failure of ${\bf d}$ to act via the Leibniz rule involves terms of the structural form $\eta^{MN}\partial_MF_1 \partial_N F_2$
and is due to $\Delta:=\eta^{MN}\partial_M\partial_N = \B-\bar{\B}$ being a second-order differential operator. 
This failure is cured by imposing the so-called strong constraint $\Delta=0$ in the sense of double field theory \cite{Hull:2009mi,Siegel:1993th,Hohm:2010pp}, 
i.e., $\Delta$ annihilates 
all fields and all their products. An obvious solution to this constraint is to identify $x=\bar{x}$. 
(One may also attempt to define a `weakly constrained' version following the recent progress in \cite{Bonezzi:2023ced,Bonezzi:2023lkx}, 
but we will not do so here.)

Decomposing the 2-form $\Psi_{MN}$ into 3D objects  one obtains $e_{\mu\bar{\nu}}$, the generalized 
metric fluctuation of double field theory \cite{Hohm:2010pp}, and two 2-forms. 
While to the best of our knowledge this (topological) 3D gravity model has not 
 been explored independently, we will point out a curious relation to Kodaira-Spencer gravity, a topological 
 gravity theory in 6D with the complex structure deformation of a Hermitian manifold as fundamental field \cite{Bershadsky:1993cx,Gopakumar:1998vy}. 
 The relation of the  field equations (\ref{PSIEQINTRO}) to the Kodaira-Spencer equations is as follows: 
 Truncating  the two 2-forms, which is a consistent truncation in the technical sense that all solutions 
 of the truncated theory  uplift to solutions of the full theory,  one is left with \textit{two}  
 equations for $e_{\mu\bar{\nu}}$, reflecting the fact that the theory is non-Lagrangian. Picking one of the two 
 equations only (which of course is no longer a consistent truncation) and rewriting it in terms of complex coordinates 
 of $\mathbb{C}^3$ one obtains the Kodaira-Spencer equations on a 3D real slice of $\mathbb{C}^3$.  
 Once we restrict to one of the equations one may relax the strong constraint as the Kodaira-Spencer
 theory is consistent in 6D thanks to its invariance under only `holomorphic' 3D diffeomorphisms. 
 We will show that the full 6D Kodaira-Spencer gravity can instead be obtained as a consistent truncation 
 of  a certain `chiral' double copy. 
 
The remainder of this paper  is organized as follows. In sec.~2 we review the kinematic BV$^{\B}$ algebra 
 of Chern-Simons theory. We then work out the double copied gravity theory and give the double field theory 
 formulation with 6D covariance, which is diffeomorphism invariant and local but non-Lagrangian. 
 In sec.~3 we show that a gauge invariant action can be written at the cost of abandoning locality. 
 In sec.~4 we rewrite the theory in terms of complex coordinates of 
 $\mathbb{C}^3$ and describe  the relation to Kodaira-Spencer gravity, which is a consistent truncation 
 of a chiral double copy. 
 We close with a summary and 
 outlook in sec.~5.

\section{Double Copy of Chern-Simons Theory}

\subsection{Kinematic algebra of Chern-Simons theory }

We begin by recalling the kinematic algebra of Chern-Simons theory. The kinematic algebra  is defined on the de Rham complex 
of an arbitrary 3-manifold. 
This chain complex  is 
the graded vector space $\cK=\bigoplus_{p=0}^3K_p$, 
with $K_p$ the space of $p$-forms and  the differential given by the de Rham differential  $d$ satisfying $d^2=0$:
\begin{equation}\label{Kdiagram}
\begin{tikzcd}[row sep=2mm]
K_0\arrow{r}{d}&K_1\arrow{r}{d}&K_2\arrow{r}{d}&K_3  \\
\lambda&A&\cE&\cN 
\end{tikzcd}
\end{equation}
Here we also indicated the field theory interpretations in the second line 
(that will emerge after taking the tensor product with the color Lie algebra $\mathfrak{g})$. Specifically, 
the gauge parameter $\lambda$ is a zero-form, the field $A$ is  a one-form, the `equation of motion' is a two-form, and 
the  `Noether identity' $\cN$ is a three-form.

This chain complex carries a graded commutative and associative  algebra structure, given by the familiar wedge 
product of differential forms. We  use the standard normalization
\begin{equation}
\omega_p=\frac{1}{p!}\,\omega_{\mu_1\cdots\mu_p}(x)\,\theta^{\mu_1}\cdots\theta^{\mu_p}\;,   \end{equation}
where we identified the basis one-forms with odd variables  $\theta^\mu$ of degree $+1$:  $dx^\mu\simeq \theta^\mu$.  
Thus, we can think of differential forms as functions $\omega(x,\theta)$, $\eta(x,\theta)$, etc., of  even coordinates $x^{\mu}$ 
and odd coordinates $\theta^{\mu}$. The wedge product is then encoded  in the ordinary point-wise product 
of functions, which is graded commutative and associative. Specifically, this defines a differential graded (dg) commutative associative algebra (dgca) 
(the strict case of a $C_{\infty}$ algebra) 
with differential $m_1$ of degree $+1$ and a product $m_2$ of degree $0$, given by 
\begin{equation}
m_1 \equiv d \equiv \theta^\mu\del_\mu\;,   \qquad
  m_2(\omega ,\eta)\equiv \omega \cdot \eta \;.    
\end{equation}
This form makes it clear  that $m_1=d$ is a first-order differential operator and hence obeys the Leibniz rule with respect to $m_2$. 
  
While the above `kinematic' dgca  exists for any topological 3-manifold, the full kinematic algebra, denoted BV$^{\B}$, 
requires more structure, as given by a choice of metric. Given such a choice of metric, which for simplicity we take to be flat and Euclidean, 
we have the Hodge star operation defined by
\begin{equation}
\star\,\omega_p=\frac{1}{p!(3-p)!}\,\epsilon_{\mu_1\cdots\mu_{3-p}\nu_1\cdots\nu_p}\,\omega^{\nu_1\cdots\nu_p}\,\theta^{\mu_1}\cdots\theta^{\mu_{3-p}}\;,  
\end{equation}
satisfying $\star^2=1$,  
where the metric is used to raise and lower  indices.  
Furthermore, we have the adjoint or divergence operator $d^\dagger$ acting on a $p-$form as 
\begin{equation}\label{ddagger}
d^\dagger \ :=  \ (-1)^{p+1}\star d\,\star \ \equiv \ 
\frac{\partial^2}{\partial x_{\mu} \partial\theta^{\mu}} \;, 
\end{equation}
where again the metric is used to lower the index on $x^{\mu}$. 
Note that $d^\dagger$ defines a second differential, satisfying $(d^\dagger)^2=0$, that is of opposite degree to $d$. 
Moreover, $d$ and $d^\dagger$ satisfy the familiar relation 
\begin{equation}\label{dddaggeranti} 
d\,d^\dagger+d^\dagger\,d=\B \equiv \partial^{\mu}\partial_{\mu} \;.    
\end{equation}
Notice that the second form of $d^\dagger$ in (\ref{ddagger}) makes it clear  that it is a second-order operator on the associative product of graded functions (a.k.a.~the wedge product of forms). 

In line with the general literature we also denote $b\equiv d^\dagger$, so that (\ref{ddagger}) reads 
 \begin{equation}\label{goodbNot} 
  b=\del^\mu\cD_\mu\;, \;\;  {\rm where} \quad  \cD_\mu:=\frac{\del}{\del\theta^\mu}\;, 
 \end{equation} 
and (\ref{dddaggeranti}) becomes 
the graded commutator relation  $[m_1,b]=\B$. In addition to the graded commutative associative product $m_2$ there is 
a derived (`kinematic') Lie bracket, defined by the \textit{failure of $b$ to act as a derivation on $m_2$}. 
This so-called antibracket can be written as the graded commutator  $b_2:=[b,m_2]$, for 
which a brief computation with (\ref{goodbNot})  gives 
\begin{equation}\label{SchoutenNijenhuisBracket} 
b_2(\omega_1,\omega_2)=\cD_\mu\omega_1\del^\mu\omega_2+(-1)^{\omega_1\omega_2}\cD_\mu\omega_2\del^\mu\omega_1  \;. 
\end{equation}
This  is a graded Lie bracket, which generalizes the familiar Poisson bracket to the graded setting and is equivalent to the Schouten-Nijenhuis bracket on polyvector fields \cite{Witten:1990wb} (see also \cite{Borsten:2022vtg} 
and sec.~2 of \cite{Bonezzi:2022bse}). Specifically, given two polyvector fields $\Pi_{1}=\tfrac{1}{p!}\, \Pi_{1}^{\mu_{1}\ldots\mu_{p}}\, \del_{\mu_{1}}\wedge\ldots\wedge \del_{\mu_{p}}$ and $\Pi_{2}=\tfrac{1}{q!}\, \Pi_{2}^{\mu_{1}\ldots\mu_{q}}\, \del_{\mu_{1}}\wedge\ldots\wedge \del_{\mu_{q}}$  the Schouten-Nijenhuis bracket acts as 
\begin{equation}
\big[\Pi_{1},\Pi_{2}\big]_{\rm{SN}} = \nabla\cdot (\Pi_{1}\wedge \Pi_{2})-\nabla\cdot \Pi_{1}\wedge \Pi_{2}-(-1)^{p}\, \Pi_{1}\wedge \nabla\cdot \Pi_{2}\;,
\end{equation}
where $\nabla\cdot$ is the covariant divergence of polyvector fields. Since we are working on flat space, the covariant divergence is simply 
\begin{equation}
(\nabla\cdot \Pi)^{\mu_{1}\ldots \mu_{p-1}} =\del_{\nu} \Pi^{\nu\mu_{1}\ldots \mu_{p-1}}\,. 
\end{equation}
One can straightforwardly identify polyvector fields and differential forms, for which  the Schouten-Nijenhuis bracket and the bracket $b_{2}$ in \eqref{SchoutenNijenhuisBracket} are equivalent. 

There is also a Poisson compatibility relation between  $b_2$ and $m_2$. 
While  $b_2$ is a graded Lie bracket, $(m_1,b_2)$  fail to define  a differential graded Lie algebra (dgLa or a strict $L_{\infty}$ algebra) 
due to the box obstruction for the Leibniz relation following from $[m_1,b]=\B$: 
\begin{equation}
[m_1,b_2]=[\B,m_2]\;.    
\end{equation}
In contrast, $(b,b_2)$ define a dgLa since $b$ acts via the Leibniz rule on $b_2$, as one may quickly verify.  
With these relations, the data $(m_1,b,m_2,b_2)$ together define a  BV$^{\B}$ algebra \cite{Reiterer:2019dys}.

We finally point out that the integration of differential forms equips ${\cal K}$ with an inner product. 
Specifically, integration of top forms coincides with the integration of graded functions:
\begin{equation}
\int \omega_3=\int d^3\theta\,d^3x\,\omega_3(x,\theta)\;,\quad\int d^3\theta\,\theta^\mu\theta^\nu\theta^\rho:=\epsilon^{\mu\nu\rho}\;,   
\end{equation}
which provides a degree $-3$ inner product on $\cK$:
\begin{equation}
\big\l \omega,\eta\big\r:=\int d^3\theta\,d^3x\, \omega(x,\theta)\,\eta(x,\theta)\equiv\int \omega\wedge \eta\;.    
\end{equation}
The pairing is non-vanishing only between $K_1$ and $K_2$ and between $K_0$ and $K_3$. This encodes  the usual pairings between fields and equations and between gauge parameters and Noether identities. The 
operator $\star$ realizes  the inner product isomorphisms $K_1\simeq K_2$ and $K_0\simeq K_3$.

\subsection{The double copy complex}

We will now realize an exact double copy of Chern-Simons theory. This  is possible since  the product $m_2$ (pointwise product of graded functions) and the bracket $b_2$ (Schouten-Nijenhuis bracket) are part of a \textit{strict}  BV$^\B$ algebra on ${\cal K}$. The double copy is then defined on ${\cal K}\otimes \bar{\cal K}$, which carries a BV$^{\Delta}$ algebra, where $\Delta=\B-\bar\B$ \cite{Bonezzi:2023lkx}. 
In order to define a consistent field theory on the tensor product ${\cal K}\otimes \bar{\cal K}$, we impose the strong constraint $\Delta=0$ or $\B=\bar\B$ on all elements $\Psi\in  {\cal K}\otimes \bar{\cal K}$ and their products, which leads to a (strict) $L_{\infty}$ algebra underlying a non-Lagrangian gravity theory.

First of all, the double copy space is given by $\cX=\cK\otimes\bar\cK$. All elements of $\cX$ are $(p,q)-$forms on the doubled space $\mathbb{R}^3\times\mathbb{R}^3$, in the sense that 
\begin{equation}
\Omega_{p,q}=\frac{1}{p!q!}\,\Omega_{\mu_1\cdots\mu_p\bar\nu_1\cdots\bar\nu_q}(x,\bar x)\,\theta^{\mu_1}\cdots\theta^{\mu_p}\,\bar\theta^{\bar\nu_1}\cdots\bar\theta^{\nu_q}\;,  
\end{equation}
with the degree in $\cX$ given by total form degree: $|\Omega_{p,q}|=p+q$. Integration of top forms extends naturally to the doubled space:
\begin{equation}\label{doubleint}
\int \Omega_{3,3}=\int d^3\bar\theta\,d^3\theta\,d^3x\,d^3\bar x\,\Omega_{3,3}(x,\bar x, \theta, \bar\theta)\;,\quad \int d^3\bar\theta \, d^3\theta\,\theta^\mu\theta^\nu\theta^\rho\,\bar\theta^{\bar\mu}\bar\theta^{\bar\nu}\bar\theta^{\bar\rho}:=\epsilon^{\mu\nu\rho}\epsilon^{\bar\mu\bar\nu\bar\rho}\;.  
\end{equation}

The total  space $\cX$  decomposes as follows:
\begin{equation}\label{Xdiagram}
\begin{tikzcd}[row sep=2mm]
X_{0}\arrow{r}{{\bf d}}&X_{1}\arrow{r}{{\bf d}}&X_2\arrow{r}{{\bf d}}&X_3\arrow{r}{{\bf d}}&X_4\arrow{r}{{\bf d}}&X_5\arrow{r}{{\bf d}}&X_6\\ 
\chi_{0,0}& \begin{array}{c} \lambda_{1,0}\\\bar\lambda_{0,1}
\end{array}&\begin{array}{c} C_{2,0}\\ e_{1,1}\\\bar C_{0,2}
\end{array}& \begin{array}{c}\cE_{3,0}\\ \cE_{2,1}\\ \bar\cE_{1,2}\\ \bar\cE_{0,3}
\end{array}& \begin{array}{c}\cN_{3,1}\\ \cN_{2,2}\\ \bar\cN_{1,3}
\end{array}& \begin{array}{c} \cR_{3,2}\\ \bar\cR_{2,3}
\end{array}&\cV_{3,3}
\end{tikzcd}   
\end{equation}
where the differential is ${\bf d}=d+\bar d$.
Above we denoted the $(p,q)-$form degrees by subscripts, and aligned vertically the elements according to the twist $p-q$.
The integration \eqref{doubleint}
induces a pairing on $\cX$ given by
\begin{equation}\label{Xpairing}
\big\l \Omega,H\big\r:=\int d^3\bar\theta\,d^3\theta\,d^3x\,d^3\bar x\, \Omega(x,\bar x,\theta,\bar\theta)\,H(x,\bar x,\theta,\bar\theta)\;,    
\end{equation}
which, for bi-forms $\Omega_{p,q}$ and $H_{r,s}$, is non-vanishing only if $p+r=q+s=3$. 
We give a field theory interpretation to the chain complex \eqref{Xdiagram} by identifying fields as elements of $X_2$. This choice is due to the fact that the metric fluctuation is expected to arise from the `tensor product of two gauge fields'. More precisely, given the Chern-Simons complex \eqref{Kdiagram} the spin two fluctuation resides in $K_1\otimes\bar K_1$. Following this interpretation gauge parameters are elements of $X_1$ and field equations live in $X_3$, while elements in higher degree correspond to a cascade of Noether and Noether-for-Noether identities. The degree convention differs from the standard $L_\infty$ one (where fields are in degree zero), but for differential forms the form degree is more natural.
From the explicit form \eqref{Xdiagram} of the complex, it is clear that the field theory described by the $L_\infty$ algebra on $\cX$ is non-Lagrangian, at least if one insists in identifying fields in $X_2$ and field  equations in $X_3$, 
for then there are more field equations than fields.

Keeping this standard interpretation, we begin by working out the linear theory. 
The fields living in $X_2$ can be parameterized as 
\begin{equation}\label{psi}
\Psi=e+C+\bar C=e_{\mu\bar\nu}\,\theta^\mu\bar\theta^{\bar\nu}-\tfrac12\,C_{\mu\nu}\,\theta^{\mu}\theta^{\nu}+\tfrac12\,\bar C_{\bar\mu\bar\nu}\,\bar\theta^{\bar\mu}\bar\theta^{\bar\nu}\;, 
\end{equation}
while the gauge parameters living in $X_{1}$ can be written as 
\begin{equation}
\Lambda=\lambda+\bar\lambda=-\lambda_\mu\,\theta^\mu+\bar\lambda_{\bar\mu}\,\bar\theta^{\bar\mu}\;. 
\end{equation}
The linear gauge transformations $\delta\Psi={\bf d}\Lambda$ then read 
\begin{align}\label{gauge}
\delta e&=d\bar\lambda+\bar d\lambda\qquad\quad \longrightarrow & \delta e_{\mu\bar\nu}&=\del_\mu\bar\lambda_{\bar\nu}+\bar\del_{\bar\nu}\lambda_\mu\;,\\  
\delta C&=d\lambda\qquad\qquad\quad\;\longrightarrow &\delta C_{\mu\nu}&=\del_\mu\lambda_\nu-\del_\nu\lambda_\mu\;,\\
\delta \bar  C&=\bar d\bar\lambda \qquad\qquad\quad\;\longrightarrow&\delta\bar C_{\bar\mu\bar\nu}&=\bar\del_{\bar\mu}\bar\lambda_{\bar\nu}-\bar\del_{\bar\nu}\bar\lambda_{\bar\mu}\;.   
\end{align}
The above gauge transformations are reducible, with trivial parameters given 
 by $\lambda_\mu^{\rm triv.}=-\del_\mu\chi$ and $\bar\lambda^{\rm triv.}_{\bar\mu}=\bar\del_{\bar\mu}\chi$. The reducibility parameter $\chi$ is the single element of $X_0$.
The gauge invariant field equations ${\bf d}\Psi=0$ split similarly into $(3,0)$, $(2,1)$, $(1,2)$ and $(0,3)$ components: 
\begin{align}\label{1stordereoms}
dC&=0\qquad\qquad\longrightarrow & \del_{\mu}C_{\nu\rho}+\del_{\nu}C_{\rho\mu}+\del_{\rho}C_{\mu\nu}&=0\;,\nonumber\\
de+\bar dC&=0 \qquad\qquad\longrightarrow &\del_\mu e_{\nu\bar\rho}-\del_\nu e_{\mu\bar\rho}-\bar\del_{\bar\rho}C_{\mu\nu}&=0\;,\nonumber\\
\bar de+ d\bar C&=0 \qquad\qquad\longrightarrow &\bar\del_{\bar\mu} e_{\rho\bar\nu}-\bar\del_{\bar\nu} e_{\rho\bar\mu}-\del_{\rho}\bar C_{\bar\mu\bar\nu}&=0\;,\nonumber\\
\bar d\bar C&=0\qquad\qquad\longrightarrow & \bar\del_{\bar\mu}\bar C_{\bar\nu\bar\rho}+\bar\del_{\bar\nu}\bar C_{\bar\rho\bar\mu}+\bar\del_{\bar\rho}\bar C_{\bar\mu\bar\nu}&=0\;.
\end{align}

In previous works on the double copy of Yang-Mills theory \cite{Bonezzi:2022bse,Bonezzi:2022yuh}, the gravity theory is defined on a subspace of $\cK\otimes\bar\cK$. This subspace, besides imposing the `weak constraint' $\B=\bar \B$, is defined to be ${\rm ker}\,b^-$, 
where $b^-:=b-\bar b$,  and in the present context is given by
\begin{equation}\label{b-}
b^-=d^\dagger-\bar d^\dagger\;,\qquad (b^-)^2=0\;,\qquad {\bf d}\,b^-+b^-{\bf d}=\B-\bar\B=0\;.    
\end{equation}
For now the equations \eqref{1stordereoms} are totally unconstrained. We have neither imposed  $b^-=0$ nor $\B-\bar{\B}=0$ on the elements of $\cX$, 
but let us next explore what consequences these constraints would have. 
For gauge parameters and fields, the $b^-$ constraint would  result in:
\begin{equation}\label{b-constraint}
\begin{split}
\del\cdot\lambda+\bar\del\cdot\bar\lambda&=0\;,\\
\del^\rho e_{\rho\bar\mu}-\bar\del^{\bar\rho}\bar C_{\bar\rho\bar\mu}&=0\;,\\
\bar\del^{\bar\rho}e_{\mu\bar\rho}-\del^\rho C_{\rho\mu}&=0\;,
\end{split}    
\end{equation}
yielding a constrained gauge symmetry. 
Upon imposing these constraints the field equations \eqref{1stordereoms} imply as 
integrability condition a gauge fixed version of the DFT equations in 3D.
To see this we take the divergence of \eqref{1stordereoms}, use (\ref{b-constraint}) and the weak constraint 
$\B=\bar{\B}$ to obtain 
\begin{equation}\label{2ndorderdiveom}
\begin{split}
\B e_{\mu\bar\nu}-\del_\mu\del\cdot e_{\bar\nu}-\bar\del_{\bar\nu}\bar\del\cdot e_\mu&=0\;,\\
\B C_{\mu\nu}-2\,\del_{[\mu}\del\cdot C_{\cdot\nu]}&=0\;,\\
\B \bar C_{\bar\mu\bar\nu}-2\,\bar\del_{[\bar\mu}\bar\del\cdot\bar C_{\cdot\bar\nu]}&=0\;. 
\end{split}    
\end{equation}
Notice that $\del\cdot\bar\del\cdot e=0$ is implied by taking a divergence of the constraint \eqref{b-constraint}. The $e_{\mu\bar\nu}$ equation in \eqref{2ndorderdiveom} coincides with a gauge fixed form of the DFT equation in 3D, 
where the DFT dilaton is set to zero \cite{Hull:2009mi}, 
which indeed leaves residual gauge transformations with $\del\cdot\lambda+\bar\del\cdot\bar\lambda=0$.  

Let us remark that the $b^-$ constraint is not needed in order to have a consistent double copy. Imposing $b^-=0$ is equivalent to a partial gauge fixing and subsequent solution  of some components of the unconstrained field equations \eqref{1stordereoms}.
We thus conclude that the first order equations \eqref{1stordereoms} imply the standard equations for a topological graviton and three two-forms ($C$, $\bar C$ and the $B-$field). 
We have also performed  a lightcone analysis to show that the first order system has no propagating degrees of freedom.

\subsection{Non-linear double copy}\label{nonlindc}

We now turn to the non-linear structure on the double copied space $\cK\otimes\bar \cK$. The 
general double copy prescription defines the differential ${\bf d}$ and bracket $B_2$ by  \cite{Bonezzi:2022bse} 
\begin{equation}\label{dB2}
\begin{split}
{\bf d}&=m_1\otimes1+1 \otimes\bar m_1\;,\\
B_2&=b_2\otimes\bar m_2-m_2\otimes\bar b_2\;,
\end{split}    
\end{equation}
where the bar denotes the BV$^{\bar \B}$ maps associated to $\bar\cK$. The space $\cK\otimes\bar\cK$ is also endowed with the $b^-$ operator discussed above and an associative product $M_2=m_2\otimes\bar m_2$. The latter allows one to rewrite the bracket in \eqref{dB2} as $B_2=[b^-,M_2]$, showing that it also has the form of a BV antibracket. Using the BV$^\B$ relations, including $[m_1,b_2]=[\B,m_2]$, it is straightforward to verify that $({\bf d}, b^-, M_2, B_2)$ define a BV$^{\Delta}$ algebra, where $\Delta=\B-\bar{\B}$.  
Since we need a genuine $L_{\infty}$ algebra to define a field theory 
on $\cK\otimes\bar \cK$, we restrict to a subspace by imposing the `strong constraint' 
that $\Delta=0$ or $\B=\bar{\B}$, 
for instance by identifying $x$ coordinates with $\bar{x}$ coordinates. 
Imposing the strong constraint, the space $\mathcal{K}\otimes\bar{\mathcal{K}}$ equipped with $({\bf{d}}, B_{2})$ defines a strict $L_{\infty}$ algebra (dgLa) that describes a gravity theory. 

In the following we give a more explicit form of the dgLa given by $({\bf d}, B_2)$. 
To this end it is convenient to introduce a covariant notation for the doubled space: We define coordinates in $\mathbb{R}^6$ as $x^M:=(x^\mu,\bar x^{\bar\mu})$, together with the odd elements $\theta^M=(\theta^\mu,\bar\theta^{\bar\mu})$. Arbitrary elements of $\cX$ are differential forms in $\mathbb{R}^6$, which we do not need to split  into  3D  components. 
The differential ${\bf d}$ is of course the de~Rham differential in 6D: 
\begin{equation}
{\bf d} =\theta^M\del_M \;,   
\end{equation}
while the gauge parameter and the field are generic one- and two-forms respectively:
\begin{equation}
\Lambda=\Lambda_M \theta^M\;,\quad\Psi=\tfrac12\,\Psi_{MN}\,\theta^M\theta^N\;.    
\end{equation}
The free theory describes elements of the de Rham cohomology in form degree two:
\begin{equation}
{\bf d}\Psi=0\;,\quad\delta\Psi={\bf d}\Lambda\;.    
\end{equation}
In order to describe the nonlinear theory, we shall introduce the $O(3,3)$ metric 
\begin{equation}\label{etaMN}
\eta_{MN}=\bpm\delta_{\mu\nu}&0\\0&-\delta_{\bar\mu\bar\nu}\epm   \;,
\end{equation}
and its inverse $\eta^{MN}$. In this section we will raise indices, when convenient,  with $\eta^{MN}$.

Denoting $\cD_M=\frac{\del}{\del\theta^M}$, the $b^-$ operator is given by $b^-=\del^M\cD_M$. The two-bracket of general  forms $F$ and $G$ (viewed  as graded functions of $x^M$ and $\theta^M$) is thus given by the failure of $\partial^M{\cal D}_M$ to act via the Leibniz rule on the pointwise product (the 6D wedge product) 
and is hence the 6D version of the bracket (\ref{SchoutenNijenhuisBracket}): 
\begin{equation}\label{6DSchouten} 
B_2(F,G)=\eta^{MN}\Big(\cD_M F\,\del_NG+(-1)^{FG}\cD_M G\,\del_NF\Big)\;.   
\end{equation}
This defines the non-linear field equations in 
the Maurer-Cartan form:
\begin{equation}\label{MC}
\cE:={\bf d}\Psi+\tfrac12\, B_2(\Psi,\Psi) =  0\;. 
\end{equation}
This equation  is gauge covariant under 
 \begin{equation} 
 \delta\Psi=D\Lambda:={\bf d}\Lambda+B_2(\Psi,\Lambda)\;, 
 \end{equation}  
using  the strong constraint in the form $\partial^MA\,\partial_MB=0$ for arbitrary $A, B$: 
\begin{equation}
\delta_\Lambda\cE=D^2\Lambda=B_2(\Lambda,\cE)+2\,\del^M\Psi\,\del_M\Lambda =B_2(\Lambda,\cE) \;.     
\end{equation}
Note  that while (\ref{6DSchouten}) defines a genuine graded 
Lie bracket on unconstrained forms in 6D, the differential acts only via the Leibniz rule after imposing  the strong constraint. 

\section{Non local Lagrangian double copy}

The nonlinear theory presented in the previous section is non-Lagrangian, as the number of equations of motion differs from the number of field components. In this section we are first going to impose the $b^-$ constraint (which, as we have discussed, entails a partial gauge fixing and solving field equations). We will then show that it is possible to construct a non local inner product, and thus an action, if one assumes that the $\mathbb{R}^3$ Laplacian $\B$ can be inverted.

\subsection{Imposing the $b^-$ constraint}

In the 6D covariant formulation, elements of the chain complex $\cX$ are differential forms of arbitrary rank in $\mathbb{R}^6$ subject to the  constraint $\B-\bar \B=\eta^{MN}\del_M\del_N=0$, written in terms of the $O(3,3)$ invariant metric $\eta_{MN}$. We also recall that the $b^-$ operator takes the form 
\begin{equation}
b^-=d^\dagger-\bar d^\dagger=\eta^{MN}\del_M\frac{\del}{\del\theta^N}\equiv\del^M\cD_M\;, 
\end{equation}
which is the 6D  divergence operator constructed from $\eta^{MN}$. We now subject all elements of $\cX$ to the $b^-$ constraint, i.e.~we consider only forms $F$ obeying $\del^M\cD_M F=0$, meaning that they are transverse with respect to the $O(3,3)$ metric. For the gauge parameter one-form $\Lambda$ and the two-form field $\Psi$ this results in the constraints \eqref{b-constraint} discussed previously. 
On this subspace the bracket $B_2$ reduces to
\begin{equation}
B_2(F,G)=b^-\big(FG\big)\;,\quad\forall\;F,G\in {\rm ker}\,b^-\;,
\end{equation}
where $FG$ is the point-wise product of graded functions encoding the 6D wedge product of forms.
For fields $\Psi\in{\rm ker}\,b^-$ the Maurer-Cartan equation reduces to
\begin{equation}
\cE={\bf d}\Psi+\tfrac12\,b^-(\Psi^2)\;,    
\end{equation}
and obeys $b^-\cE=0$ thanks to \eqref{b-}. Similarly, the gauge transformation with constrained parameter satisfying  $b^-\Lambda=0$ takes the form $\delta\Psi={\bf d}\Lambda+b^-(\Psi\Lambda)$.

\subsection{Non local $c^-$ operator and Lagrangian}

In certain cases, such as in the local double copy of Yang-Mills theory of \cite{Bonezzi:2022yuh,Bonezzi:2022bse}, it is possible to define an operator $c^-$ obeying $(c^-)^2=0$ and $b^-c^-+c^-b^-=1$. The chain complex $\cX$ of the double copy then splits into ${\rm ker}\,b^-\oplus{\rm im}\,c^-$, using that $b^-c^-$ and $c^-b^-$ are projectors. One can then define an inner product on ${\rm ker}\,b^-\subset\cX$, via a $c^-$ insertion\footnote{This is analogous to the $c_0^-$ insertion used to define the inner product in closed string field theory.} in the natural pairing inherited from $\cK\otimes\bar\cK$ (c.f.~\eqref{doubleint}). Given such an inner product, the known examples suggest that the corresponding double copy ought to be Lagrangian \cite{Bonezzi:2022yuh,Borsten:2023ned}.

In the present case of Chern-Simons theory there seem to be no local candidate for such a $c^-$ operator. However, if one assumes $\B$ to be invertible, a non-local $c^-$ can be defined by $c^-=\frac{m_1}{\B}-\frac{\bar m_1}{\bar\B}$. 
Let us then  assume that the 3D Laplacian $\B$ can be inverted and set 
\begin{equation}
c^-:=\frac12\,\left(\frac{d}{\B}-\frac{\bar d}{\bar\B}\right)=\frac{1}{2\,\B}\,(d-\bar d)\;,   
\end{equation}
where $d$ and $\bar d$ are the two copies of the 3D de Rham differential, and we used the strong constraint $\B=\bar\B$. This operator indeed obeys
\begin{equation}
(c^-)^2=0\;,\qquad {\bf d}c^-+c^-{\bf d}=0\;,\qquad b^-c^-+c^-b^-=1 \;,  
\end{equation}
which can be easily seen upon using ${\bf d}=d+\bar d$ and $b^-=d^\dagger-\bar d^\dagger$. The graded commutator $[b^-,c^-]=1$ ensures that any element in ker$\,b^-$ is $b^-$ exact, since for any $F$ obeying $b^-F=0$ one has 
\begin{equation}\label{b-exact}
F=(b^-c^-+c^-b^-)F=b^-c^-F\;.    
\end{equation}
We now use this to prove that the Maurer-Cartan equation $\cE=0$ is equivalent to the non local equation $c^-\cE=0$. The implication $\cE=0\;\Rightarrow\;c^-\cE=0$ is obvious. In the other direction we apply $b^-$ to $c^-\cE$:
\begin{equation}
c^-\cE=0\;\Rightarrow\;0=b^-c^-\cE=\cE\;,    
\end{equation}
where we used that $\cE=b^-c^-\cE$, since $\cE\in{\rm ker\,b^-}$.

The next step is to show that $c^-\cE=0$ is a Lagrangian equation. To this end we recall the pairing \eqref{Xpairing}, written here in 6D form:
\begin{equation}\label{pairing6D}
\l F,G\r=\int d^6x\, d^6\theta\,F(x,\theta)\,G(x,\theta)\;,\qquad \l F,G\r=(-1)^{FG}\l G,F\r\;,
\end{equation}
which picks the $(3,3)$ form component of $FG$, thanks to \eqref{doubleint}.
We claim that the equation $c^-\cE=c^-{\bf d}\Psi+\tfrac12\,c^-b^-(\Psi^2)=0    
$
results from varying the action\footnote{In $L_\infty$ terms this is the standard action obtained from an $L_\infty$ inner product defined by $\l F,G\r_{L}:=\l F,c^-G\r$.}
\begin{equation}\label{MC action}
S=\frac12\,\big\l\Psi,c^-{\bf d}\Psi\big\r+\frac{1}{3!}\,\big\l\Psi,\Psi^2\big\r \;.   
\end{equation}
To prove this we need the integration by parts relations
\begin{equation}\label{IBP}
\begin{split}
\l {\bf d}F,G\r&=(-1)^{F+1}\l F,{\bf d}G\r\;,\qquad \l c^-F,G\r=(-1)^{F+1}\l F,c^-G\r\;,\\
\l b^-F,G\r&=(-1)^{F}\l F,b^-G\r \;,    
\end{split}   
\end{equation}
which are derived by expressing ${\bf d}$ and $c^-$ in terms of $d$ and $\bar d$, together with $b^-=\del^M\frac{\del}{\del\theta^M}$.
We now compute the variation of \eqref{MC action}:
\begin{equation}
\delta S=\frac12\,\big\l\delta\Psi,c^-{\bf d}\Psi\big\r+\frac12\,\big\l\Psi,c^-{\bf d}\delta\Psi\big\r+\frac{1}{2}\,\big\l\delta\Psi,\Psi^2\big\r\;,       
\end{equation}
where for the last term we used that $\l\Psi,\Psi^2\r=\int d^6x\,d^6\theta\,\Psi^3$ is manifestly symmetric. The second term can be rewritten using \eqref{IBP} and graded symmetry of the pairing as
\begin{equation}
\begin{split}
\big\l\Psi,c^-{\bf d}\delta\Psi\big\r&=\big\l c^-{\bf d}\delta\Psi,\Psi\big\r =\big\l{\bf d}\delta\Psi,c^-\Psi\big\r\\
&=-\big\l\delta\Psi,{\bf d}c^-\Psi\big\r=\big\l\delta\Psi,c^-{\bf d}\Psi\big\r\;,
\end{split}    
\end{equation}
thus yielding $\delta S=\l\delta\Psi,c^-{\bf d}\Psi+\tfrac12\,\Psi^2\r       
$. One has to be careful in reading off the equation of motion from the variation, since $\Psi$ is constrained to obey $b^-\Psi=0$. Using \eqref{b-exact}, one has $\Psi=b^-c^-\Psi$. Taking the variation with the explicit projector on ker$\,b^-$ we find
\begin{equation}
\begin{split}
\delta S&=\big\l b^-c^-\delta\Psi,c^-{\bf d}\Psi+\tfrac12\,\Psi^2\big\r=-\big\l c^-\delta\Psi,b^-c^-{\bf d}\Psi+\tfrac12\,b^-(\Psi^2)\big\r\\
&=\big\l \delta\Psi,c^-b^-c^-{\bf d}\Psi+\tfrac12\,c^-b^-(\Psi^2)\big\r=\big\l \delta\Psi,c^-{\bf d}\Psi+\tfrac12\,c^-b^-(\Psi^2)\big\r\;,
\end{split}    
\end{equation}
where in the last equality we used $b^-c^-{\bf d}\Psi={\bf d}b^-c^-\Psi={\bf d}\Psi$. This proves that the action \eqref{MC action} yields $c^-\cE=0$ as field equations which are, in turn, equivalent to the local first order equations $\cE=0$.
Using the above formula for the general variation one proves that the action \eqref{MC action} is gauge invariant under $\delta\Psi={\bf d}\Lambda+b^-(\Psi\Lambda)$, with $b^-\Lambda=0$.

We conclude this section by writing explicitly the action \eqref{MC action} in terms of the component fields $e_{\mu\bar\nu}$, 
$C_{\mu\nu}$, and $\bar C_{\bar\mu\bar\nu}$. 
Using the decomposition \eqref{psi} of $\Psi$ and the explicit form \eqref{doubleint} for the pairing we find
\begin{equation}\label{Action explicit}
 S=\int d^3xd^3\bar x\,\epsilon^{\mu\nu\rho}\epsilon^{\bar\mu\bar\nu\bar\rho}\Big[-\frac12\,\del_\mu e_{\nu\bar\rho}\,\frac{1}{\B}\bar\del_{\bar\mu}e_{\rho\bar\nu}+\frac14\,\del_\mu C_{\nu\rho}\frac{1}{\B}\bar\del_{\bar\mu}\bar C_{\bar\nu\bar\rho}-\frac16\,e_{\mu\bar\mu}e_{\nu\bar\nu}e_{\rho\bar\rho}-\frac14\,e_{\mu\bar\mu}C_{\nu\rho}\bar C_{\bar\nu\bar\rho}\Big] \;.  
\end{equation}
This action makes contact with the ones constructed in \cite{Ben-Shahar:2021zww,Szabo:2023cmv,Borsten:2023ned} as double copies of Chern-Simons theory. In particular, the superfield formulation of the action in \cite{Ben-Shahar:2021zww} should also feature the two-form fields $C_{\mu\nu}$ and $\bar C_{\bar\mu\bar\nu}$, although it is not stated explicitly. Let us emphasize that the action \eqref{Action explicit} is gauge invariant, and it yields field equations that are equivalent to local ones. Similar non-local actions have also appeared  in topological string theory \cite{Witten:1992fb, Bershadsky:1994sr}.

\section{Relation to Kodaira-Spencer Gravity}

In this section we point out a relation between the local double copy theory constructed in section \ref{nonlindc} and the topological  theory 
given by 
Kodaira-Spencer gravity, which describes a complex structure deformation of a Hermitian manifold. To that end, we first rewrite the theory 
in complex coordinates and discuss how a truncation of that theory yields  Kodaira-Spencer gravity on a  3D subspace. Finally, we show how 
to obtain Kodaira-Spencer gravity directly by means of a chiral double copy. 

\subsection{Complex coordinates} 

As mentioned above,  the chain complex $\cK\otimes\bar\cK$ is identical to the de Rham complex on $\mathbb{R}^6$. The two copies of the Euclidean metric on $\mathbb{R}^3$ can be combined into a background ``generalized metric'' $H_{MN}$, which is just the flat Euclidean metric of $\mathbb{R}^6$, and the $O(3,3)$ metric $\eta_{MN}$: 
\begin{equation}
H_{MN}=\bpm \delta_{\mu\nu}&0\\0&\delta_{\bar\mu\bar\nu}\epm\;,\quad  \eta_{MN}=\bpm \delta_{\mu\nu}&0\\0&-\delta_{\bar\mu\bar\nu}\epm\;.
\end{equation}
The metric $H_{MN}$ does not look like the one of a Hermitian manifold, which has only mixed components in holomorphic coordinates. On top of that, the coordinates $x^\mu$ and $\bar x^{\bar\mu}$ are real. To remedy this we define a set of complex coordinates given by
\begin{equation}\label{z defined}
\begin{split}
z^1&:=\tfrac{1}{\sqrt 2}(x+i\bar x)\;,\quad z^2:=\tfrac{1}{\sqrt 2}(y+i\bar y)\;,\quad z^3:=\tfrac{1}{\sqrt 2}(z+i\bar z)\;,\\
\bar z^{\bar 1}&:=\tfrac{1}{\sqrt 2}(x-i\bar x)\;,\quad \bar z^{\bar 2}:=\tfrac{1}{\sqrt 2}(y-i\bar y)\;,\quad \bar z^{\bar 3}:=\tfrac{1}{\sqrt 2}(z-i\bar z)\;,
\end{split}
\end{equation}
which we collectively denote as $(z^a,\bar z^{\bar a})$. In these coordinates, the metric $H_{MN}$ becomes
\begin{equation}
ds_H^2=dx^2+dy^2+dz^2+d\bar x^2+d\bar y^2+d\bar z^2=2\,\delta_{a\bar b}\,dz^ad\bar z^{\bar b}  \;,  
\end{equation}
which is the expected form of a Hermitian flat metric in $\mathbb{C}^3$.
On the other hand, the $O(3,3)$ metric becomes
\begin{equation}
ds^2_\eta=dx^2+dy^2+dz^2-d\bar x^2-d\bar y^2-d\bar z^2=\delta_{ab}\,dz^adz^b+\delta_{\bar a \bar b}\,d\bar z^{\bar a}d\bar z^{\bar b}\;,   
\end{equation}
so that we have
\begin{equation}
H_{AB}=\bpm0&\delta_{a\bar b}\\\delta_{\bar a b}&0\epm \;,\quad \eta_{AB}=\bpm\delta_{ab}&0\\0&\delta_{\bar a\bar b}\epm \;,   
\end{equation}
in coordinates $z^A:=(z^a,\bar z^{\bar a})$. We similarly redefine the one-form basis by introducing $(\theta^a,\bar\theta^{\bar a})$ in the same fashion: $\theta^1:=\tfrac{1}{\sqrt 2}(\theta^x+i\bar \theta^{\bar x})$ and so on. The 6D volume form can also be factorized into its (anti)holomorphic components: $\epsilon_{abc}\,\epsilon_{\bar a\bar b\bar c}\,$.

With these redefinitions, the elements of $\cK\otimes\bar \cK$ are given by all differential forms in $\mathbb{C}^3$:
\begin{equation}
\cK\otimes\bar\cK=\bigoplus_{p,q=0}^3\Omega^{p,q}(\mathbb{C}^3)\;,   
\end{equation}
where now we split degrees according to the number of $(\theta^a, \bar\theta^{\bar a})$. The differential is the de Rham differential in six dimensions, which splits into its Dolbeault components:
\begin{equation}
 {\bf d} =\theta^a\del_a+\bar\theta^{\bar a}\bar\del_{\bar a}=\del+\bar\del\;. \end{equation}
Redefining the fields according to the new holomorphic degree we write
\begin{equation}
\Psi=h_{1,1}+B_{2,0}+\bar B_{0,2}\;.    
\end{equation}
Their linearized Maurer-Cartan equations  read 
\begin{equation}
\begin{split}
\del B&=0\;,\quad \bar\del\bar B=0\;,\\
\del h+\bar\del B&=0\;,\quad \bar\del h+\del\bar B=0\;,
\end{split}    
\end{equation}
which are  invariant under the gauge transformations
\begin{equation}
\delta h=\del\bar\lambda+\bar\del\lambda\;,\quad\delta B=\del\lambda\;,\quad\delta\bar B=\bar \del\bar\lambda\;.
\end{equation}

If we want to continue with the exact quadratic Maurer-Cartan  equation, we have to make sense of the $b$ operators and the 
strong constraint. 
Let us start with the $b$ operators. Using the redefinitions we have
\begin{equation}
b^+=H^{AB}\del_A\frac{\del}{\del\theta^B}=\del^\dagger+\bar\del^\dagger\;,\qquad b^-=\eta^{AB}\del_A\frac{\del}{\del\theta^B}=\delta+\bar\delta\;,    
\end{equation}
where we have the adjoints of the Dolbeault operators:
\begin{equation}
\del^\dagger=\delta^{a\bar b}\bar\del_{\bar b}\frac{\del}{\del\theta^a}\;,\quad \bar\del^\dagger=\delta^{a\bar b}\del_{a}\frac{\del}{\del\bar\theta^{\bar b}}\;, 
\end{equation}
which obey
\begin{equation}
\begin{split}
\del\del^\dagger+\del^\dagger\del=\tfrac12\,\B=\delta^{a\bar b}\del_a\bar\del_{\bar b}\;,   
\end{split}    
\end{equation}
with the same commutator for the barred ones, and 
\begin{equation}
\delta:=\delta^{ab}\del_{a}\frac{\del}{\del\theta^b}\;,\qquad \bar\delta:=\delta^{\bar a\bar b}\bar\del_{\bar a}\frac{\del}{\del\bar\theta^{\bar b}}\;, 
\end{equation}
which employ 
the chiral Kronecker deltas. 
Note that here we encounter an unusual feature: 
the $b^-$ operator involves new structures, which are the Kronecker deltas that, although  present in $\mathbb{C}^3$, do not exist for generic  Hermitian manifolds.  We should think of this structure associated to $\mathbb{C}^3$ as an auxiliary  structure, 
on par with the auxiliary metric introduced in 3D Chern-Simons theory.  
Using this structure, the (weak or strong) constraint reads 
\begin{equation}
\Delta:=\eta^{AB}\del_A\del_B=\delta^{ab}\del_a\del_b+\delta^{\bar a \bar b}\bar\del_{\bar a}\bar\del_{\bar b} = 0  \;.  
\end{equation}

\subsection*{The interacting theory}

Let us compute explicitly what the above  field equations and gauge symmetries look like.
The Lie bracket $B_2\equiv[\cdot,\cdot]$ has the same six-dimensional form as before:
\begin{equation}
\big[F,G\big]=\eta^{AB}\Big(\cD_A F\,\del_BG+(-1)^{FG}\cD_A G\,\del_BF\Big)\;.   
\end{equation}
The Maurer-Cartan equation for the two-form $\Psi=\frac12\,\theta^A\theta^B\,\psi_{AB}$ reads
\begin{equation}
\del_{[A}\psi_{BC]}-\psi_{[A}{}^D\del_{|D|}\psi_{BC]}
= 0\;,    
\end{equation}
where the indices are raised with $\eta^{AB}$. The above equation is covariant under the gauge transformations
\begin{equation}
\delta\psi_{AB}=2\,\del_{[A}\lambda_{B]}-2\,\psi_{[A}{}^C\del_{|C|}\lambda_{B]}+\lambda^C\del_C\psi_{AB}\;,    
\end{equation}
provided the strong constraint is obeyed, since the variation of the field equation reads
\begin{equation}\label{obstruction}
\delta_\lambda\cE_{ABC}= [\Lambda,\cE]_{ABC}+6\,\del^D\psi_{[AB}\,\del_{|D|}\lambda_{C]}\;.    
\end{equation}
Splitting into holomorphic and antiholomorphic components, we have four field equations:
\begin{equation}\label{holo eom}
\begin{split}
2\,\del_{[a} h_{b]\bar c}-2\,h_{[a}{}^{\bar d}\bar\del_{|\bar d|}h_{b]\bar c}
+\bar\del_{\bar c}B_{ab}-2\,B_{[a}{}^d\del_{|d|}h_{b]\bar c}+h^d{}_{\bar c}\del_dB_{ab}-\bar B_{\bar c}{}^{\bar d}\bar\del_{\bar d}B_{ab}&= 0\;,\\
-2\,\bar \del_{[\bar a} h_{|c|\bar b]}-2\, h^{d}{}_{[\bar a}\del_{|d|}h_{|c|\bar b]}
+\del_{c}\bar B_{\bar a\bar b}+2\,\bar B_{[\bar a}{}^{\bar d}\bar \del_{|\bar d|}h_{|c|\bar b]}-h_{c}{}^{\bar d}\bar \del_{\bar d}\bar B_{\bar a\bar b}-B_{c}{}^{d}\del_{d}\bar B_{\bar a\bar b}&= 0\;,\\
3\,\del_{[a} B_{bc]}-3\,h_{[a}{}^{\bar d}\bar\del_{|\bar d|}B_{bc]}-3\,B_{[a}{}^d\del_{|d|}B_{bc]}&= 0\;,\\
3\,\bar \del_{[\bar a} \bar B_{\bar b\bar c]}+3\,h^{d}{}_{[\bar a}\del_{|d|}\bar B_{\bar b\bar c]}-3\,\bar B_{[\bar a}{}^{\bar d}\bar \del_{|\bar d|}\bar B_{\bar b\bar c]}&= 0\;,
\end{split}    
\end{equation}
where we used the definition $h_{\bar b a}:=-h_{a\bar b}$.
Splitting the gauge transformations
we obtain
\begin{equation}\label{holo transf}
\begin{split}
\delta h_{a\bar b}&=\del_a\bar\lambda_{\bar b}-\bar\del_{\bar b}\lambda_{a}-h_a{}^{\bar c}\bar\del_{\bar c}\bar\lambda_{\bar b}-h^{c}{}_{\bar b}\del_{c}\lambda_{a}+\bar B_{\bar b}{}^{\bar c}\bar\del_{\bar c}\lambda_{a}-B_a{}^c\del_c\bar\lambda_{\bar b}+\bar\lambda^{\bar c}\bar\del_{\bar c}h_{a\bar b}+\lambda^{c}\del_{c}h_{a\bar b}  \;,\\
\delta B_{ab}&= 2\,\del_{[a}\lambda_{b]} -2\,h_{[a}{}^{\bar c}\bar\del_{\bar c}\lambda_{b]}-2\,B_{[a}{}^c\del_c\lambda_{b]}+\lambda^c\del_cB_{ab}+\bar\lambda^{\bar c}\bar\del_{\bar c}B_{ab}\;, \\
\delta \bar B_{\bar a\bar b}&= 2\,\bar \del_{[\bar a}\bar \lambda_{\bar b]} +2\,h^{c}{}_{[\bar a}\del_{c}\bar \lambda_{\bar b]}-2\,\bar B_{[\bar a}{}^{\bar c}\bar \del_{\bar c}\bar \lambda_{\bar b]}+\lambda^c\del_c\bar B_{\bar a\bar b}+\bar\lambda^{\bar c}\bar\del_{\bar c}\bar B_{\bar a\bar b}\;.
\end{split}    
\end{equation}

Let us first try to interpret the standard solution to the strong constraint, which in holomorphic coordinates  reads
\begin{equation}
\delta^{ab}\del_aF\,\del_bG+\delta^{\bar a\bar b}\bar\del_{\bar a}F\,\bar\del_{\bar b}G=0 \;. 
\end{equation}
The supergravity solution in the $(x^\mu,\bar x^{\bar\mu})$ coordinates is $x^\mu=\bar x^{\bar\mu}$, thus also identifying the corresponding derivatives.
Looking at the definition \eqref{z defined} of the complex coordinates, this amounts to setting
$z^a=i\,\bar z^{\bar a}$. This is a three dimensional real slice of $\mathbb{C}^3$ given by a diagonal hyperplane.

Since the strong constraint is required to get rid of the obstruction \eqref{obstruction} in the gauge variation of the Maurer-Cartan 
equation, let us see in more detail which component fields and gauge parameters are involved. Splitting \eqref{obstruction} in (anti)holomorpic components we have
\begin{equation}\label{holo obstruction}
\begin{split}
\delta_\lambda\cE_{abc}&=[\Lambda,\cE]_{abc}+6\,\del^DB_{[ab}\,\del_{|D|} \lambda_{c]}\;,\\
\delta_\lambda\cE_{ab\bar c}&= [\Lambda,\cE]_{ab\bar c}+2\,\del^DB_{ab}\,\del_D\bar\lambda_{\bar c}-4\,\del^Dh_{[a|\bar c }\,\del_{D|}
\lambda_{b]} \;,
\end{split}    
\end{equation}
with the other two components obtained by swapping barred and unbarred indices and fields.
From the above expression one can see that the strong constraint would not be required if we truncated $B_{ab}=0$ and $\lambda_a=0$, which can also be checked by inspection of \eqref{holo eom} and \eqref{holo transf}. The reason this is not consistent in our model is of course that we have two more equations, obtained from \eqref{holo obstruction} by (un-)barring the indices. Demanding consistency of the other equations without the strong constraint would require $\bar B_{\bar a\bar b}=0$ and $\bar\lambda_{\bar a}=0$, thus leaving no gauge symmetry whatsoever.

Nevertheless, it is still interesting to see what happens if we do forget the other two equations, and set
$B_{ab}=0$ and $\lambda_a=0$.
The system \eqref{holo eom} with gauge transformations \eqref{holo transf} then reduces to
\begin{equation}\label{KSfirst}
\begin{split}
\del_a h_b{}^{\bar c}-\del_b h_a{}^{\bar c}-h_a{}^{\bar d}\,\bar\del_{\bar d}h_b{}^{\bar c}+h_b{}^{\bar d}\,\bar\del_{\bar d}h_a{}^{\bar c}=0\;,\\
\delta h_a{}^{\bar b}=\del_a\bar\lambda^{\bar b}-h_a{}^{\bar c}\,\bar\del_{\bar c}\bar\lambda^{\bar b}+\bar\lambda^{\bar c}\,\bar\del_{\bar c}h_a{}^{\bar b}  \;,
\end{split}    
\end{equation}
where we raised the antiholomorphic index with $\delta^{\bar a\bar b}$. The equations in \eqref{KSfirst} are the Kodaira-Spencer equation and gauge transformation for $h$, upon  interpreting $h_a{}^{\bar b}$ and $\bar\lambda^{\bar a}$ as holomorphic one- and zero-forms, respectively, taking values in antiholomorphic vector fields:
\begin{equation}\label{handl}
h:=dz^a\,h_a{}^{\bar b}\,\bar\del_{\bar b}\;,\quad \lambda:=\bar\lambda^{\bar a}\,\bar\del_{\bar a}\;.  
\end{equation}
Of course, as derived the theory is still subject to the strong constraint and hence three-dimensional, but in this form the theory is actually 
gauge invariant without the constraint and can hence be lifted to 6D.

\subsection{Kodaira-Spencer gravity  as a chiral double copy} 

We now turn to a chiral double copy construction which yields a genuine six-dimensional field theory containing the Kodaira-Spencer equation \cite{Bershadsky:1993cx,Gopakumar:1998vy}. 
To this end we recall that, in the standard prescription used in the previous sections, the differential ${\bf d}$ and bracket $B_2$ on $\cK\otimes\bar \cK$ are given by
\begin{equation}
\begin{split}
{\bf d}&=m_1\otimes1+1\otimes\bar m_1\;,\\
B_{2}&=b_2\otimes\bar m_2-m_2\otimes\bar b_2\;,
\end{split}    
\end{equation}
where $(m_1,m_2,b_2)$ and their barred counterparts are the BV$^\B$ maps in $\cK$ and $\bar\cK$, respectively. The need for the strong constraint originates  from the failure of $m_1$ to be a derivation of the bracket $b_2$, since it implies that ${\bf d}$ fails to be a derivation of $B_2$ unless $\B=\bar\B$. 

For the following construction we will treat $\cK$ and $\bar\cK$ differently: we still view $\cK$ as the space of forms in $\mathbb{R}^3$ carrying the dgca $(m_1=d,m_2=\wedge)$. On the other hand, we  treat $\bar\cK$ as the space of polyvectors in $\mathbb{R}^3$, forming a graded Lie algebra with bracket $\bar b_2$, the Schouten-Nijenhuis bracket. On $\cX=\cK\otimes\bar\cK$ 
one then has  a differential graded Lie algebra given by
\begin{equation}
{\bf d}=m_1\otimes 1\;,\quad B_{2}=m_2\otimes\bar b_2\;,     \end{equation}
which is defined on $\mathbb{R}^6$, with coordinates $z^A=(z^a,\bar z^{\bar a})$. The elements of $\cX$ are polyvector-valued forms of the following type:
\begin{equation}
    \omega_p^q=\frac{1}{p!q!}\, \omega_{a_{1}\ldots a_{p}}{}^{\bar b_{1}\ldots \bar b_{q}}(z,\bar z)\, \theta^{a_{1}}\cdots \theta^{a_{p}}\bar\theta_{\bar b_{1}}\cdots \bar\theta_{\bar b_{q}}\;.
\end{equation}
The degree on $\cX$ is the sum of the form and polyvector degrees, so that $|\omega_p^q|=p+q$. On $\mathbb{R}^6$ we can choose the following complex structure:
\begin{equation}
J^A{}_B=i\bpm\delta^a{}_b&0\\
0&-\delta^{\bar a}{}_{\bar b}\epm\;.    
\end{equation}
In the coordinates $z^A=(z^a,\bar z^{\bar a})$ this identifies $\theta^a$ as the basis of holomorphic one-forms and $\bar\theta_{\bar a}$ as the basis of anti-holomorphic vector fields on $\mathbb{C}^3$.
With this complex structure at hand, the differential ${\bf d}\equiv \del=\theta^a\del_a$ is the holomorphic Dolbeault differential, while the two-bracket $B_{2}$ acts  as
\begin{equation}
B_{2}(\omega_{1},\omega_{2})=\frac{\del \omega_{1}}{\del\bar\theta_{\bar a}}\bar\del_{\bar a}\omega_{2}+(-1)^{\omega_{1}\omega_{2}} \frac{\del \omega_{2}}{\del\bar\theta_{\bar a}}\bar\del_{\bar a}\omega_{1}\;,
\end{equation}
where the degree in the exponent is the sum of the form degree and the polyvector field degree.

The gauge parameters $\Lambda$ and fields $\Psi$ split as 
\begin{equation}
\begin{split}
    \Lambda &= \lambda_{a}\theta^{a}+\bar\xi^{\bar a}\bar\theta_{\bar a}\;,\\
    \Psi &= h_{a}{}^{\bar b}\theta^{a}\bar\theta_{\bar b}+\tfrac{1}{2}\, B_{ab}\theta^{a}\theta^{b}+\tfrac{1}{2}\, \Pi^{\bar a\bar b}\bar\theta_{\bar a}\bar\theta_{\bar b}\;.
\end{split}
\end{equation}
Gauge parameters thus consist of a holomorphic one-form $\lambda_a$ and an anti-holomorphic vector field $\bar\xi^{\bar a}$. Fields contain a vector-valued one-form $h_a{}^{\bar b}$, a two-form $B_{ab}$ and a bivector $\Pi^{\bar a\bar b}$\footnote{This field content coincides with the one alluded to by Witten in the context of topological string theory \cite{Witten:1992fb}.}. In order to interpret this set of parameters and fields we compute the gauge algebra, which is given by the bracket $B_2(\Lambda_1,\Lambda_2)$. Its one-form and vector components read
\begin{equation}
\begin{split}
B_{2}(\Lambda_{1},\Lambda_{2})_{a}&=\bar \xi_{1}^{\bar b}\, \bar\del_{\bar b}\lambda_{2\, a}-\bar\xi^{\bar b}_{2}\, \bar\del_{\bar b}\lambda_{1\, a}=\cL_{\bar\xi_1}\lambda_{2\,a}-\cL_{\bar\xi_2}\lambda_{1\,a}\,,  \\
B_{2}(\Lambda_{1},\Lambda_{2})^{\bar a}&=\bar\xi^{\bar b}_{1}\, \bar\del_{\bar b} \bar\xi^{\bar a}_{2}-\bar\xi^{\bar b}_{2}\, \bar\del_{\bar b} \bar\xi^{\bar a}_{1}=\cL_{\bar\xi_1}\bar\xi_2^{\bar a}\;,
\end{split}
\end{equation}
with $\cL$ denoting the Lie derivative.
This shows that the gauge transformations are the semi-direct sum of abelian holomorphic one-form transformations and anti-holomorphic diffeomorphisms.
The transformations of the fields themselves are given by $\delta\Psi=\del\Psi+B_2(\Psi,\Lambda)$, yielding
\begin{equation}
\begin{split}
    \delta h_{a}{}^{\bar b}&=\del_{a}\bar\xi^{\bar b}+\bar\xi^{\bar c}\, \bar\del_{\bar c}h_{a}{}^{\bar b}-\del_{\bar c}\bar\xi^{\bar b}\, h_{a}{}^{\bar c}+\Pi^{\bar b\bar c}\, \bar\del_{\bar c}\lambda_{a}=\del_{a}\bar\xi^{\bar b}+\cL_{\bar\xi}h_a{}^{\bar b}+\Pi^{\bar b\bar c}\, \bar\del_{\bar c}\lambda_{a}\;,\\
    \delta B_{ab}&=2\, \big(\del_{[a} -h_{[a}{}^{\bar c}\bar\del_{\bar c}\big)\lambda_{b]}+\bar\xi^{\bar c}\, \bar\del_{\bar c}B_{ab}=2\,\del^h_{[a}\lambda_{b]}+\cL_{\bar\xi}B_{ab}\;,\\
    \delta \Pi^{\bar a\bar b}&= \bar \xi^{\bar c}\, \bar\del_{\bar c}\Pi^{\bar a\bar b}-\bar\del_{\bar c}\bar\xi^{\bar a}\, \Pi^{\bar c\bar b}-\bar\del_{\bar c}\bar\xi^{\bar b}\, \Pi^{\bar a\bar c}=\cL_{\bar\xi}\Pi^{\bar a\bar b}\;,
\end{split}
\end{equation}
where we defined the twisted derivative $\del_a^h:=\del_a-h_a{}^{\bar b}\bar\del_{\bar b}$. If one interprets $\theta^a\del_a^h$ as a deformed Dolbeault operator, the field $h_a{}^{\bar b}$ is a deformation of the complex structure. The gauge transformation of $h_a{}^{\bar b}$ under diffeomorphisms is the expected one, but the additional contribution from the one-form $\lambda_a$ is exotic.
The field equations $\del\Psi+\frac12 B_2(\Psi,\Psi)=0$ decompose as
\begin{equation}
\begin{split}
2\, \del_{[a}h_{b]}{}^{\bar c}-2\, h_{[a}{}^{\bar d}\, \bar\del_{|\bar d|}h_{b]}{}^{\bar c}-\Pi^{\bar c\bar d}\, \bar\del_{\bar d}B_{ab}&=0\; ,\\
3\, (\del_{[a}-h_{[a}{}^{\bar c}\bar \del_{|\bar c|})B_{bc]}&=0\;,\\
\del_{a}\Pi^{\bar b\bar c}-h_{a}{}^{\bar d}\, \bar\del_{\bar d}\Pi^{\bar b\bar c}-2\, \Pi^{\bar d[\bar b}\, \bar\del_{\bar d}h_{a}{}^{\bar c]}&=0\;,\\
3\, \Pi^{\bar d[\bar a}\, \bar\del_{\bar d}\Pi^{\bar b\bar c]}&=0\;.
\end{split}
\end{equation}
Interestingly, the field equation for $\Pi^{\bar a\bar b}$ is the condition for it to be a Poisson bivector compatible, in a suitable sense, with the deformation $h_a{}^{\bar b}$. In addition, the deformed Dolbeault operator
\begin{equation}
\del_h:=dz^a(\del_a-h_a{}^{\bar b}\bar\del_{\bar b}) \;,   
\end{equation}
is not nilpotent, with the two-form $B_{ab}$ sourcing the curvature for $\del_h^2$. The two-form itself has vanishing (twisted) field strength on-shell.

The above is a gauge invariant dynamical system, which is novel to the best of our knowledge,  that generalizes the Kodaira-Spencer equation and its gauge symmetries. The Kodaira-Spencer equation is contained as a consistent truncation, setting  $\Pi^{\bar a\bar b}=0$  and 
then removing  the two-form by means of the $\lambda$ gauge symmetry. The resulting theory is defined by 
\begin{equation}\label{KSsecond}
\begin{split}
\del_a h_b{}^{\bar c}-\del_b h_a{}^{\bar c}-h_a{}^{\bar d}\,\bar\del_{\bar d}h_b{}^{\bar c}+h_b{}^{\bar d}\,\bar\del_{\bar d}h_a{}^{\bar c}=0\;,\\
\delta h_a{}^{\bar b}=\del_a\bar\lambda^{\bar b}-h_a{}^{\bar c}\,\bar\del_{\bar c}\bar\lambda^{\bar b}+\bar\lambda^{\bar c}\,\bar\del_{\bar c}h_a{}^{\bar b}  \;,
\end{split}    
\end{equation}
which is Kodaira-Spencer gravity.\footnote{See \cite{Mason:2007ct, Borsten:2023paw} for similar results on holomorphic Chern-Simons theories in twistor space, which are related to self-dual Yang-Mills theory and self-dual supergravities.}

\section{Summary and Outlook} 

In this paper we have constructed the double copy of 3D Chern-Simons theory, employing the same algebraic double copy recipe 
previously applied to pure Yang-Mills theory to obtain ${\cal N}=0$ supergravity in a double field theory formulation. 
In this we believe to have given the first example of an explicit, local and gauge invariant double copy construction  that is exact, 
albeit being non-Lagrangian. An action can, however, be constructed upon partial gauge fixing and giving up locality. 
This result is hence   an important ingredient of the larger research program of double copying   general field theories. 
 
 Chern-Simons theory may also turn out to be an interesting toy model  
 due to some similarities  with the self-dual sector of Yang-Mills theory \cite{Bonezzi:2023pox}: in both the double copied gravity theory features  a generalized metric fluctuation and two 2-forms. 
 The question arises whether the 
 non-linear field equations of self-dual gravity or even full gravity may be recovered  as integrability conditions, 
 which likely would require a novel procedure to eliminate the extra 2-form fields. 
Another important question is whether there might be a novel perspective to eventually obtain a local and gauge invariant action. 

Finally, one of the core motivations for this program has always been  to establish a precise 
correspondence between classical  solutions 
of  gauge theory and gravity. Since here the double copied 3D gravity theory can be written down 
explicitly in terms of the kinematic ingredients of 3D Chern-Simons theory, this example may be a promising starting point 
to attempt the double copy of solutions.


 \section*{Acknowledgments} 

We thank Maor Ben-Shahar, Christoph Chiaffrino, Henrik Johansson, Michael Reiterer, Davide Scazzuso and Barton Zwiebach for discussions. 

\noindent
This work is funded   by the European Research Council (ERC) under the European Union's Horizon 2020 research and innovation programme (grant agreement No 771862)
and by the Deutsche Forschungsgemeinschaft (DFG, German Research Foundation), ``Rethinking Quantum Field Theory", Projektnummer 417533893/GRK2575. The work of R.B. is funded by the Deutsche Forschungsgemeinschaft (DFG, German Research
Foundation)–Projektnummer 524744955.


\begin{thebibliography}{99}



\bibitem{Bern:2008qj}
Z.~Bern, J.~J.~M.~Carrasco and H.~Johansson,
``New Relations for Gauge-Theory Amplitudes,''
Phys. Rev. D \textbf{78}, 085011 (2008)
[arXiv:0805.3993 [hep-ph]].

\bibitem{Bern:2019prr}
Z.~Bern, J.~J.~Carrasco, M.~Chiodaroli, H.~Johansson and R.~Roiban,
``The Duality Between Color and Kinematics and its Applications,''
[arXiv:1909.01358 [hep-th]].


\bibitem{Monteiro:2014cda}
R.~Monteiro, D.~O'Connell, and C.~D. White, ``{Black holes and the double
  copy}'', \href{http://dx.doi.org/10.1007/JHEP12(2014)056}{{\em JHEP} {\bf 12}
  (2014)  056}, \href{http://arxiv.org/abs/1410.0239}{{\tt arXiv:1410.0239
  [hep-th]}}.


\bibitem{Luna:2015paa}
A.~Luna, R.~Monteiro, D.~O'Connell, and C.~D. White, ``{The classical double
  copy for Taub\textendash{}NUT spacetime}'',
  \href{http://dx.doi.org/10.1016/j.physletb.2015.09.021}{{\em Phys. Lett. B}
  {\bf 750} (2015)  272--277}, \href{http://arxiv.org/abs/1507.01869}{{\tt
  arXiv:1507.01869 [hep-th]}}.

\bibitem{Luna:2016hge}
A.~Luna, R.~Monteiro, I.~Nicholson, A.~Ochirov, D.~O'Connell, N.~Westerberg,
  and C.~D. White, ``{Perturbative spacetimes from Yang-Mills theory}'',
  \href{http://dx.doi.org/10.1007/JHEP04(2017)069}{{\em JHEP} {\bf 04} (2017)
  069}, \href{http://arxiv.org/abs/1611.07508}{{\tt arXiv:1611.07508
  [hep-th]}}.


\bibitem{Luna:2018dpt}
A.~Luna, R.~Monteiro, I.~Nicholson, and D.~O'Connell, ``{Type D Spacetimes and
  the Weyl Double Copy}'',
  \href{http://dx.doi.org/10.1088/1361-6382/ab03e6}{{\em Class. Quant. Grav.}
  {\bf 36} (2019)  065003}, \href{http://arxiv.org/abs/1810.08183}{{\tt
  arXiv:1810.08183 [hep-th]}}.


\bibitem{Kim:2019jwm}
K.~Kim, K.~Lee, R.~Monteiro, I.~Nicholson, and D.~Peinador~Veiga, ``{The
  Classical Double Copy of a Point Charge}'',
  \href{http://dx.doi.org/10.1007/JHEP02(2020)046}{{\em JHEP} {\bf 02} (2020)
  046}, \href{http://arxiv.org/abs/1912.02177}{{\tt arXiv:1912.02177
  [hep-th]}}.


\bibitem{Monteiro:2021ztt}
R.~Monteiro, S.~Nagy, D.~O'Connell, D.~Peinador~Veiga, and M.~Sergola, ``{NS-NS
  spacetimes from amplitudes}'',
  \href{http://dx.doi.org/10.1007/JHEP06(2022)021}{{\em JHEP} {\bf 06} (2022)
  021}, \href{http://arxiv.org/abs/2112.08336}{{\tt arXiv:2112.08336
  [hep-th]}}.


\bibitem{Monteiro:2020plf}
R.~Monteiro, D.~O'Connell, D.~Peinador~Veiga, and M.~Sergola, ``{Classical
  solutions and their double copy in split signature}'',
  \href{http://dx.doi.org/10.1007/JHEP05(2021)268}{{\em JHEP} {\bf 05} (2021)
  268}, \href{http://arxiv.org/abs/2012.11190}{{\tt arXiv:2012.11190
  [hep-th]}}.

\bibitem{Bern:2007hh}
Z.~Bern, J.~J. Carrasco, L.~J. Dixon, H.~Johansson, D.~A. Kosower, and
  R.~Roiban, ``{Three-Loop Superfiniteness of N=8 Supergravity}'',
  \href{http://dx.doi.org/10.1103/PhysRevLett.98.161303}{{\em Phys. Rev. Lett.}
  {\bf 98} (2007)  161303}, \href{http://arxiv.org/abs/hep-th/0702112}{{\tt
  arXiv:hep-th/0702112}}.

\bibitem{Bern:2010ue}
Z.~Bern, J.~J.~M. Carrasco, and H.~Johansson, ``{Perturbative Quantum Gravity
  as a Double Copy of Gauge Theory}'',
  \href{http://dx.doi.org/10.1103/PhysRevLett.105.061602}{{\em Phys. Rev.
  Lett.} {\bf 105} (2010)  061602}, \href{http://arxiv.org/abs/1004.0476}{{\tt
  arXiv:1004.0476 [hep-th]}}.


\bibitem{Carrasco:2011mn}
J.~J.~M. Carrasco and H.~Johansson, ``{Five-Point Amplitudes in N=4
  Super-Yang-Mills Theory and N=8 Supergravity}'',
  \href{http://dx.doi.org/10.1103/PhysRevD.85.025006}{{\em Phys. Rev. D} {\bf
  85} (2012)  025006}, \href{http://arxiv.org/abs/1106.4711}{{\tt
  arXiv:1106.4711 [hep-th]}}.


\bibitem{Bjerrum-Bohr:2013iza}
N.~E.~J. Bjerrum-Bohr, T.~Dennen, R.~Monteiro, and D.~O'Connell, ``{Integrand
  Oxidation and One-Loop Colour-Dual Numerators in N=4 Gauge Theory}'',
  \href{http://dx.doi.org/10.1007/JHEP07(2013)092}{{\em JHEP} {\bf 07} (2013)
  092}, \href{http://arxiv.org/abs/1303.2913}{{\tt arXiv:1303.2913 [hep-th]}}.

\bibitem{He:2015wgf}
S.~He, R.~Monteiro, and O.~Schlotterer, ``{String-inspired BCJ numerators for
  one-loop MHV amplitudes}'',
  \href{http://dx.doi.org/10.1007/JHEP01(2016)171}{{\em JHEP} {\bf 01} (2016)
  171}, \href{http://arxiv.org/abs/1507.06288}{{\tt arXiv:1507.06288
  [hep-th]}}.


\bibitem{Bern:2018jmv}
Z.~Bern, J.~J. Carrasco, W.-M. Chen, A.~Edison, H.~Johansson,
  J.~Parra-Martinez, R.~Roiban, and M.~Zeng, ``{Ultraviolet Properties of
  $\mathcal N = 8$ Supergravity at Five Loops}'',
  \href{http://dx.doi.org/10.1103/PhysRevD.98.086021}{{\em Phys. Rev. D} {\bf
  98} (2018) no.~8, 086021}, \href{http://arxiv.org/abs/1804.09311}{{\tt
  arXiv:1804.09311 [hep-th]}}.


\bibitem{Edison:2022jln}
A.~Edison, S.~He, H.~Johansson, O.~Schlotterer, F.~Teng, and Y.~Zhang,
  ``{Perfecting one-loop BCJ numerators in SYM and supergravity}'',
  \href{http://dx.doi.org/10.1007/JHEP02(2023)164}{{\em JHEP} {\bf 02} (2023)
  164}, \href{http://arxiv.org/abs/2211.00638}{{\tt arXiv:2211.00638
  [hep-th]}}.


\bibitem{Bern:2023zkg}
Z.~Bern, J.~J.~M. Carrasco, M.~Chiodaroli, H.~Johansson, and R.~Roiban,
  ``{Supergravity amplitudes, the double copy and ultraviolet behavior}'',
  \href{http://arxiv.org/abs/2304.07392}{{\tt arXiv:2304.07392 [hep-th]}}.


\bibitem{Zwiebach:1992ie}
B.~Zwiebach,
``Closed string field theory: Quantum action and the B-V master equation,''
Nucl. Phys. B \textbf{390}, 33-152 (1993)
[arXiv:hep-th/9206084 [hep-th]].

\bibitem{Lada:1992wc}
T.~Lada and J.~Stasheff,
``Introduction to SH Lie algebras for physicists,''
Int. J. Theor. Phys. \textbf{32}, 1087-1104 (1993)
[arXiv:hep-th/9209099 [hep-th]].

\bibitem{Zeitlin:2008cc}
A.~M.~Zeitlin,
``Conformal Field Theory and Algebraic Structure of Gauge Theory,''
JHEP \textbf{03}, 056 (2010)
[arXiv:0812.1840 [hep-th]].

\bibitem{Hohm:2017pnh}
O.~Hohm and B.~Zwiebach,
``$L_{\infty}$ Algebras and Field Theory,''
Fortsch. Phys. \textbf{65}, no.3-4, 1700014 (2017)
[arXiv:1701.08824 [hep-th]].


\bibitem{Zeitlin:2009tj}
A.~M.~Zeitlin,
``Quasiclassical Lian-Zuckerman Homotopy Algebras, Courant Algebroids and Gauge Theory,''
Commun. Math. Phys. \textbf{303}, 331-359 (2011)
[arXiv:0910.3652 [math.QA]].

\bibitem{Zeitlin:2014xma}
A.~M.~Zeitlin,
``Beltrami-Courant differentials and $G_{\infty}$-algebras,''
Adv. Theor. Math. Phys. \textbf{19}, 1249-1275 (2015)
[arXiv:1404.3069 [math.QA]].

\bibitem{Borsten:2021hua}
L.~Borsten, H.~Kim, B.~Jur\v{c}o, T.~Macrelli, C.~Saemann and M.~Wolf,
``Double Copy from Homotopy Algebras,''
Fortsch. Phys. \textbf{69}, no.8-9, 2100075 (2021)
[arXiv:2102.11390 [hep-th]].


\bibitem{Reiterer:2019dys}
M.~Reiterer,
``A homotopy BV algebra for Yang-Mills and color-kinematics,''
[arXiv:1912.03110 [math-ph]].

\bibitem{Borsten:2022vtg}
L.~Borsten, B.~Jurco, H.~Kim, T.~Macrelli, C.~Saemann and M.~Wolf,
``Kinematic Lie Algebras from Twistor Spaces,''
Phys. Rev. Lett. \textbf{131}, no.4, 041603 (2023)
doi:10.1103/PhysRevLett.131.041603
[arXiv:2211.13261 [hep-th]].


\bibitem{Bonezzi:2022bse}
R.~Bonezzi, C.~Chiaffrino, F.~Diaz-Jaramillo and O.~Hohm,
``Gauge invariant double copy of Yang-Mills theory: The quartic theory,''
Phys. Rev. D \textbf{107}, no.12, 126015 (2023)
[arXiv:2212.04513 [hep-th]].

\bibitem{Batalin:1981jr}
I.~A.~Batalin and G.~A.~Vilkovisky,
``Gauge Algebra and Quantization,''
Phys. Lett. B \textbf{102}, 27-31 (1981). 

\bibitem{CarrilloVallette}
I.~Galvez-Carrillo, Imma, A.~Tonks, and B.~Vallette. 
``Homotopy Batalin-Vilkovisky algebras," Journal of Noncommutative Geometry 6.3 (2012): 539-602.


\bibitem{Hull:2009mi}
C.~Hull and B.~Zwiebach,
``Double Field Theory,''
JHEP \textbf{09}, 099 (2009)
[arXiv:0904.4664 [hep-th]].

\bibitem{Siegel:1993th}
W.~Siegel,
``Superspace duality in low-energy superstrings,''
Phys. Rev. D \textbf{48}, 2826-2837 (1993)
[arXiv:hep-th/9305073 [hep-th]].

\bibitem{Hohm:2010pp}
O.~Hohm, C.~Hull and B.~Zwiebach,
``Generalized metric formulation of double field theory,''
JHEP \textbf{08}, 008 (2010)
[arXiv:1006.4823 [hep-th]].

\bibitem{Bonezzi:2022yuh}
R.~Bonezzi, F.~Diaz-Jaramillo and O.~Hohm,
``The gauge structure of double field theory follows from Yang-Mills theory,''
Phys. Rev. D \textbf{106}, no.2, 026004 (2022)
[arXiv:2203.07397 [hep-th]].

\bibitem{Diaz-Jaramillo:2021wtl}
F.~Diaz-Jaramillo, O.~Hohm and J.~Plefka,
``Double field theory as the double copy of Yang-Mills theory,''
Phys. Rev. D \textbf{105}, no.4, 045012 (2022)
[arXiv:2109.01153 [hep-th]].


\bibitem{Ben-Shahar:2021zww}
M.~Ben-Shahar and H.~Johansson,
``Off-shell color-kinematics duality for Chern-Simons,''
JHEP \textbf{08}, 035 (2022)
[arXiv:2112.11452 [hep-th]].

\bibitem{Szabo:2023cmv}
R.~J.~Szabo and G.~Trojani,
``Homotopy Double Copy of Noncommutative Gauge Theories,''
Symmetry \textbf{15} (2023) no.8, 1543
doi:10.3390/sym15081543
[arXiv:2306.12175 [hep-th]].

\bibitem{Borsten:2023ned}
L.~Borsten, B.~Jurco, H.~Kim, T.~Macrelli, C.~Saemann and M.~Wolf,
``Double Copy from Tensor Products of Metric BV${}^{\color{gray} \blacksquare}$-algebras,''
[arXiv:2307.02563 [hep-th]].


\bibitem{Bonezzi:2023ced}
R.~Bonezzi, C.~Chiaffrino, F.~Diaz-Jaramillo and O.~Hohm,
``Weakly constrained double field theory: the quartic theory,''
[arXiv:2306.00609 [hep-th]].


\bibitem{Bonezzi:2023lkx}
R.~Bonezzi, C.~Chiaffrino, F.~Diaz-Jaramillo and O.~Hohm,
``Weakly constrained double field theory as the double copy of Yang-Mills theory,''
Phys. Rev. D \textbf{109}, no.6, 066020 (2024)
[arXiv:2309.03289 [hep-th]].




\bibitem{Bershadsky:1993cx}
M.~Bershadsky, S.~Cecotti, H.~Ooguri and C.~Vafa,
``Kodaira-Spencer theory of gravity and exact results for quantum string amplitudes,''
Commun. Math. Phys. \textbf{165} (1994), 311-428
[arXiv:hep-th/9309140 [hep-th]].

\bibitem{Gopakumar:1998vy}
R.~Gopakumar and C.~Vafa,
``Topological gravity as large N topological gauge theory,''
Adv. Theor. Math. Phys. \textbf{2} (1998), 413-442
doi:10.4310/ATMP.1998.v2.n2.a8
[arXiv:hep-th/9802016 [hep-th]].

\bibitem{Witten:1990wb}
E.~Witten,
``A Note on the Antibracket Formalism,''
Mod. Phys. Lett. A \textbf{5} (1990), 487
doi:10.1142/S0217732390000561

\bibitem{Witten:1992fb}
E.~Witten,
``Chern-Simons gauge theory as a string theory,''
Prog. Math. \textbf{133} (1995), 637-678
[arXiv:hep-th/9207094 [hep-th]].


\bibitem{Bershadsky:1994sr}
M.~Bershadsky and V.~Sadov,
``Theory of Kahler gravity,''
Int. J. Mod. Phys. A \textbf{11} (1996), 4689-4730
doi:10.1142/S0217751X96002157
[arXiv:hep-th/9410011 [hep-th]].

\bibitem{Mason:2007ct}
L.~J.~Mason and M.~Wolf,
``Twistor Actions for Self-Dual Supergravities,''
Commun. Math. Phys. \textbf{288}, 97-123 (2009)
doi:10.1007/s00220-009-0732-5
[arXiv:0706.1941 [hep-th]].

\bibitem{Borsten:2023paw}
L.~Borsten, B.~Jurco, H.~Kim, T.~Macrelli, C.~Saemann and M.~Wolf,
``Double-copying self-dual Yang-Mills theory to self-dual gravity on twistor space,''
JHEP \textbf{11}, 172 (2023)
doi:10.1007/JHEP11(2023)172
[arXiv:2307.10383 [hep-th]].


\bibitem{Bonezzi:2023pox}
R.~Bonezzi, F.~Diaz-Jaramillo and S.~Nagy,
``Gauge independent kinematic algebra of self-dual Yang-Mills theory,''
Phys. Rev. D \textbf{108} (2023) no.6, 065007
[arXiv:2306.08558 [hep-th]].

 \end{thebibliography}
\end{document}